\documentclass[12pt]{article}
\def\kakko#1{\langle #1 \rangle}
\begin{document}
\begin{flushright}
hep-ph/0103181
\end{flushright}
\vskip 0.5cm
\begin{center}
\Large\bf
Pion Production by Parametric Resonance Mechanism 
with Quantum Back Reactions
\end{center}
\vskip 1.0cm
\begin{center}
Hideaki Hiro-Oka\footnote{hiro-oka@phys.clas.kitasato-u.ac.jp}\\

{\it Institute of Physics, Kitasato University \\
1-15-1 Kitasato Sagamihara, Kanagawa 228-8555, Japan}\\
\vskip 0.5cm
Hisakazu Minakata\footnote{minakata@phys.metro-u.ac.jp}\\
{\it Department of Physics, Tokyo Metropolitan University \\
1-1 Minami-Osawa, Hachioji, Tokyo 192-0397, Japan \\ and \\
Research Center for Cosmic Neutrinos,
Institute for Cosmic Ray Research, \\
University of Tokyo, Kashiwa, Chiba 277-8582, Japan}
\end{center}
\vskip 1.5cm
\begin{abstract}
We investigate the problem of quantum back reactions due to 
particle production by the parametric resonance mechanism in 
an environment of nonequilibrium chiral phase transition.
We work with the linear sigma model and employ Boyanovsky et al.'s 
formalism to take account of back reactions under the Hartree-Fock 
approximation. 
We calculate the single pion momentum distributions and two pion 
correlations under the initial condition of thermal equilibrium 
and small amplitude sigma oscillations around a potential minimum.
We observe that the resonance peak survives under the back reaction 
which is remarkable considering the strong coupling of the sigma model.

\end{abstract}

\begin{flushleft}
{11.10.Ef,25.75.Gz,25.75.Dw}
\end{flushleft}
\clearpage

\section {Introduction}

It has been suggested that occurrence of nonequilibrium chiral
phase transition in hadronic collisions would lead to the formation
of disoriented chiral condensate (DCC), domains of coherent
pion fields \cite {DCCearly}. It then may offer a viable
explanation of the Centauro events found in cosmic ray
experiments which have anomalously large fluctuations of
neutral to charged pion ratio \cite {centauro}.

By a numerical simulation using the linear sigma model,
Rajagopal and Wilczek \cite{RW93} demonstrated that
low-momentum components of pion fields are greatly amplified,
indicating the formation of large coherent domains of pions.
A key ingredient in their simulation is the \lq\lq quench" initial
condition which models a hypothesized rapid cooling of hot
debris formed during the hadronic reactions.
The formation of large domains of pion fields was confirmed by
more realistic but still the sigma model simulation by
Asakawa, Huang, and Wang \cite {AHW95} in which the longitudinal 
expansion of hadronic blob is taken into account.

There remain, however, questions regarding the interpretation
of the simulations. Among other things, the amplification of the
low-momentum modes in the Rajagopal-Wilczek simulation lasts so
long, much longer than $1/m_{\sigma}$, the time scale of rolling
down from the top of the Mexican hat potential.
In previous works \cite{HM00,mina}, we have conjectured that the
parametric resonance mechanism might be relevant for the DCC formation. 
In simple physical terms, the parametric resonance mechanism stands 
for a mechanism of amplification of pion field fluctuations due to 
cooperative effects coupled with oscillations of background sigma 
model fields.

It has been known since long time ago that the parametric resonance 
mechanism can be relevant for cosmology by providing reheating mechanism 
during inflation \cite{brand}. It has been received renewed interests 
recently in relationship with various issues, such as the gravitino problem, 
in inflationary cosmology 
\cite{linde}\nocite{boyan3}-\cite{yoshimura}.
The parametric resonance mechanism for pion production was first 
discussed by Mr\'owczy\'nski and M\"uller 
in close analogy to the out-of-equilibrium phase transition in 
cosmology \cite {MM95}.
We have stressed in our previous paper \cite {HM00} 
that it may offer an efficient mechanism of pion production with 
potential possibility of explaining long-lasting amplification of 
low momentum modes. We also noted that it would give an alternative 
picture of large isospin fluctuations, the picture quite different 
from the conventional one based on rolling down into the
\lq\lq wrong" direction in isospin space and the subsequent relaxation
to the true vacuum. We have discussed that the discriminative
signature can be provided by the two pion correlations
in back-to-back momentum configurations \cite {HM00,HM98}.

In this paper, we investigate the effect of quantum back reactions 
onto the parametric resonance mechanism. 
%%% insert
Particle creation, when it occurs due to strong coupling with 
background field oscillations, affects the motion of the background 
fields by acting as dissipation. 
The back reaction, in turn, affects the particle production 
itself because of the damping of background oscillations.
We aim at taking account of the interplay between particle production, 
its back reactions to the background fields, and reaction back to 
the particle production in a self consistent manner. 
%%%%
Unfortunately, the effect was completely ignored in our previous 
treatment, but evaluation of its effects is indispensable for the correct 
understanding of a role played by the parametric resonance mechanism 
on DCC formation. This is our first trial toward the goal and we explore 
in this paper the effect of back reactions on single particle
momentum distributions of pions, with primary concern on its
effect on resonance peaks. We will also calculate correlation length 
of pions, one of the key issues in DCC. 

In our previous works, our formulation relied on the
approximation of small background sigma oscillations.
It may correspond to the final stage of nonequilibrium
chiral phase transition and we discussed the quantum particle
creation under the background. We have argued that in this
setting we may ignore higher order terms in the quantum pion 
and sigma fluctuations.
Due to a series of approximations, we are able to write down
explicit quantum states of pions and sigmas in the form of 
the squeezed states \cite{optics}. It enables us to compute 
explicitly two pion correlations, and the BCS type back-to-back
momentum pairing is obtained. It would give a characteristic 
signature of the parametric resonance mechanism.

In this paper, we calculate the effect of quantum back 
reactions but still within the framework of small amplitude background
sigma oscillations. Our restriction to the small-amplitude region 
is not only due to technical reasons, but also is physically motivated.
To understand the effects of quantum back reactions, it is desirable 
to compare the results with and without back reactions at the same 
initial amplitude of background $\sigma$ oscillation.
Of course, the limitation clearly implies that the meaning of our 
results must be carefully interpreted, in particular with regard to 
possible implications to experiments. 

We use the formulation by Boyanovsky et al. \cite {boyan3}
which incorporates the quantum back reactions via the Hartree-Fock
approximation. In our previous computation ignoring back reactions,
the pion momentum distributions have resonance peaks at the momenta
which are characteristic to the parametric resonance mechanism.
If observed, it would give an unambiguous confirmation of the
mechanism. However, it is possible that it can be wiped out when
the effect of quantum back reaction is taken into account. 
For related works which address DCC by the similar formalism, 
see e.g., Refs. \cite {kaiser,fred,boyan1}.

We will observe that the quantum back reactions drastically affect
low momentum components of the pion field fluctuations but {\it not}
on the resonance peaks. The peak hight is barely changed
but the peak position is moved to lower momentum.
We believe that the fact that the resonance peak survives even if 
the quantum back reactions are taken into account gives us some 
positive hints about experimental detectability of the parametric 
resonance mechanism.

Our calculation of two pion correlations results in a short 
correlation length of order 1-2 fm which is much shorter compared 
to the one expected in DCC scenario. We will briefly address its 
possible interpretation, in particular in the relationship with 
the different result obtained by Kaiser \cite {kaiser}.

In section 2, we review the Boyanovsky et al.'s formalism. 
In section 3, we calculate the single pion momentum distributions 
and two pion correlations. 
In the last section, we give concluding remarks.

\section {Quantum Back Reactions in the Hartree-Fock Approximation}

We start with the following linear sigma model Lagrangian
\begin{equation}
{\cal L}={1\over 2} \partial_\mu\phi_a \partial^\mu\phi_a
- V(\phi),\label{lag1}
\end{equation}
where
\begin{equation}
V(\phi)=\frac{\lambda}{4}(\phi_a\phi_a - v_0^2)^2+ h\sigma, \quad
\phi_a=(\sigma, \vec\pi).\label{potential}
\end{equation}
Typical values of the parameters
\begin{equation}
\lambda = 20,\quad v_0 = 90~\mbox{MeV},\quad
m_\pi=\sqrt{h\over v_0} = 140~\mbox{MeV},
\label {parameters}
\end{equation}
are employed throughout this paper, the same ones as used in our 
previous analysis. \cite {HM00}.
We rely on the formalism based on the Schr\"odinger picture
formulation of quantum field theory \cite{boyan3}.

In the following calculations, we consider the pion production
in a background sigma field which is oscillating along the
sigma direction in isospin space.
Decomposing sigma model fields into the background and fluctuations,
$\sigma=\varphi_0+\varphi$ and $\vec\pi=\vec\pi$,
we substitute them into the Lagrangian (\ref{lag1}).
We employ the Hartree-Fock approximation to take account of the back
reaction. Then, the quartic terms in the Lagrangian can be replaced 
by quadratic terms, e.g.,
$\varphi^4\rightarrow 6\kakko{\varphi^2}\varphi^2-3\kakko{\varphi^2}^2$.
The Hamiltonian is expressed as
\begin{equation}
\begin{array}{rl}
{\cal H}=&\displaystyle{1\over 2}\Pi_\varphi^2+
\displaystyle{1\over 2}\Pi_\pi^2+\displaystyle{1\over 2}(\nabla\varphi)^2
+\displaystyle{1\over 2}(\nabla\pi)^2
+\displaystyle{1\over 2}{\cal M}_\varphi^2(t)\varphi^2
+\displaystyle{1\over 2}{\cal M}_\pi^2(t)\pi^2\cr
&+\left(\lambda(3\kakko{\varphi^2}+\kakko{\pi^2})\varphi_0+
\displaystyle{\left.
{\partial V(\sigma, 0)\over{\partial\sigma}}
\right|_{\sigma=\varphi_0}}\right)\varphi
-\displaystyle{3\over 4}\lambda\left(\kakko{\varphi^2}^2+
\kakko{\pi^2}^2\right)+V(\varphi_0,0),\label{effH}
\end{array}
\end{equation}
where $\Pi$'s stand for the conjugate momenta of fields indicated as
indices. The time dependent effective masses for sigma and pion are
written, respectively, as
\begin{equation}
\begin{array}{rl}
{\cal M}_\sigma^2=&
\lambda(3\varphi_0^2-v_0^2)+3\lambda\kakko{\varphi^2},\cr
{\cal M}_\pi^2=&
\lambda(\varphi_0^2-v_0^2)+3\lambda\kakko{\pi^2}.\label{mass}
\end{array}
\end{equation}
The expectation value is evaluated by
\begin{equation}
\kakko{\varphi^2}=\mbox{\rm Tr}\left(\varphi^2\rho\right),
\end{equation}
where $\rho$ is a functional density matrix written in terms of $\varphi$.
In a self-consistent Hartree approximation,
the density matrix in sigma sector takes the Gaussian form \cite{boyan3}
\begin{equation}
\rho_\sigma[\sigma, \tilde\sigma]=\prod_k {\cal N_\varphi}\exp\left[
-{{\cal A}_\varphi\over 2}\varphi_k\varphi_{-k}
-{{\cal A}^*_\varphi\over 2}\tilde\varphi_k\tilde\varphi_{-k}
-{\cal B}_\varphi\varphi_k\tilde\varphi_{-k}
+i\Pi_{\varphi k}(\varphi_{-k}-\tilde\varphi_{-k})
\right],\label{density}
\end{equation}
where ${\cal N}_\varphi$, ${\cal A_\varphi}$, ${\cal A^*_\varphi}$,
and ${\cal B}_\varphi$ are determined so that the density matrix obeys
the functional Liouville equation
\begin{equation}
i{\partial\rho\over{\partial t}}=[{\cal H}, \rho].
\end{equation}
For instance, we have
\begin{equation}
i\dot{\cal N}_\varphi={1\over 2}{\cal N}_\varphi({\cal A}_\varphi
-{\cal A}_\varphi^*),
\end{equation}
\begin{equation}
-\dot\Pi_{\varphi 
k}=\left((3\lambda\kakko{\varphi^2}+3\lambda\kakko{\pi^2})
\varphi_0+\left.
{\partial V(\sigma,
0)\over{\partial\sigma}}\right|_{\sigma=\varphi_0}\right)
\sqrt{\Omega}\delta_{k0},\label{origin_eq_phi}
\end{equation}
where $\Omega$ is a finite volume of the system. Actually, the
equation (\ref{origin_eq_phi}) leads to the
following equation of motion of $\varphi_0$
\begin{equation}
\ddot\varphi_0+\lambda(\varphi_0^2-v_0^2)\varphi_0
+(3\lambda\kakko{\varphi^2}+3\lambda\kakko{\vec\pi^2})\varphi_0-h=0.
\label{eqphi0}
\end{equation}

As was done in the analysis by Boyanovsky et al. \cite{boyan3},
the assumption of no cross correlations between sigma and pion
fields is adopted in the derivation of the above effective
Hamiltonian (\ref{effH}). Thanks to this assumption, the density
matrices of sigma and pion sectors are factorized, allowing their
separate treatment.
Yet, the interaction between sigma and pion fields influences the
dynamics through motion of $\varphi_0$, whose behavior is strongly 
affected by quantum back reactions due to particle productions.
Among other things, it ensures the conservation of total energy
despite the factorization of sigma and pion sectors.
To briefly summarize, the effects of higher order terms of fields are
included under the Hartree-Fock approximation, and the back
reaction due to particle production is taken into account through
the mean field values evaluated with use of the density matrix.

The expectation value of sigma fluctuation is expressed as
\begin{equation}
\begin{array}{rl}
\kakko{\varphi^2}&=\displaystyle{1\over\Omega}
\sum_k\kakko{\varphi_k\varphi_{-k}}\cr
&=\displaystyle{1\over\Omega}
\sum_k {\rm Tr}\left(\varphi_k\varphi_{-k}\rho_\sigma\right).\label{VEV}
\end{array}
\end{equation}
After a short calculation, which is straightforward with the Gaussian
density
matrix, we find
\begin{equation}
\kakko{\varphi_k\varphi_{-k}}={1\over 2}\left|\psi_\varphi\right|^2
\coth\left({\beta_0\omega_\varphi(k,0)\over 2}\right)\label{one-point}
\end{equation}
under the assumption that the initial state is in thermal
equilibrium with a temperature $T_0=1/\beta_0$.
This assumption can be expressed in terms
of coefficients of the density matrix as follows:
\begin{eqnarray}
{\cal A}_\varphi(k,0) &=&
{\cal A}^*_\varphi(k,0)=
\omega_\varphi(k,0)\coth\beta_0
\omega_\varphi(k,0),\\
{\cal B}_\varphi(k,0) &=&
-\omega_\varphi(k,0)\mbox{\rm cosech}
\beta_0\omega_\varphi(k,0),\\
%\quad
{\cal N}_\varphi(k,0) &=& \left({\omega_\varphi(k,0)\over\pi}\tanh
{\beta_0\omega_\varphi(k,0)\over 2}\right)^{1/2}.
\end{eqnarray}
The function $\psi_\varphi$ defined by
\begin{equation}
-i{\dot\psi_\varphi\over\psi_\varphi}=
\mbox{\rm Re}{\cal A}_\varphi\cdot\tanh\beta_0\omega_\varphi(k,0)+
i\mbox{\rm Im}{\cal A}_\varphi\label{def_psi}
\end{equation}
obeys the equation of motion of $\varphi$,
\begin{equation}
\ddot\psi_\varphi(k,t)+
\omega_\varphi^2(k,t)\psi_\varphi(k,t)=0,
\label{eqphi}
\end{equation}
with
\begin{equation}
\omega_\varphi^2(k,t)=k^2+{\cal M}_\varphi^2(t).\label{fre_s}
\end{equation}
The initial condition for $\psi_\varphi$ is given by
\begin{equation}
\psi_\varphi(k,0)={1\over{\sqrt{\omega_\varphi(k,0)}}},\quad
\dot\psi_\varphi(k,0)=i\sqrt{\omega_\varphi(k,0)},\label{initial_s}
\end{equation}
through the initial conditions for the density matrix
under the assumption of thermal equilibrium.
The point is that the function $\psi$ satisfies the classical
equation thanks to the transformation (\ref{def_psi}),
despite that it describes the quantum system.
As one can see, the right hand side of the equation
(\ref{one-point}) contains the expectation value of $\varphi^2$
at $t=0$ through the effective mass dependence
in the frequency (\ref{fre_s}).
Hence, the initial value of $\kakko{\varphi^2}$ must be
determined self-consistently.

Let us consider an expectation value of the pion number
$\kakko{n_{\vec\pi}(k,t)}$.
A particle picture in a time dependent background, which
corresponds to the time-dependent mass (\ref{mass}), has the
similar feature as those in theories in curved space-time.
The effective mass has the similar form as the coefficient of
the quadratic term of scalar fields coupled with
the Ricci scalar \cite{linde}.
In this case, the Fock spaces chosen in both of the asymptotic
regions ($t\rightarrow \pm\infty$) belong to different Hilbert
spaces. A vacuum in one of the Fock spaces is a condensed state
of particles defined in the other Fock vacuum.
The creation and annihilation operators in each
Fock space are related via the Bogoliubov transformation.
Two different definitions of this transformation are employed in
several literatures so far
\cite{boyan3}-\nocite{STB}\nocite{kaiser}\nocite{fred}\cite{boyan1}.

In general, the Bogoliubov transformation between operators in
two Fock spaces is written as
\begin{equation}
a_k(t)=\alpha_k(t)a_{0k}+\beta_k^*(t)a_{0-k}^\dagger,\quad
a_k^\dagger(t)=\beta_k(t)a_{0-k}+\alpha_k^*(t)a_{0k}^\dagger,\label{bogo}
\end{equation}
where each of $a_k$ and $a_{0k}$ belongs to a different Fock space and
the coefficients obey $|\alpha_k(t)|^2-|\beta_k(t)|^2=1$.
From the transformation (\ref{bogo}),
the number operator $a_k^\dagger(t)a_k(t)$ is of the form
\begin{equation}
a_k^\dagger(t)a_k(t)=\left(|\alpha_k(t)|^2+
|\beta_k(t)|^2\right)a_k^\dagger a_k+
|\beta_k(t)|^2.\label{number}
\end{equation}
A possible choice of coefficients of the Bogoliubov transformation
is given by
\begin{equation}
\begin{array}{rl}
\alpha_k(t)&=\displaystyle{1\over{2\sqrt{\omega_{\vec\pi}(k,0)}}}
(i\dot\psi_{\vec\pi}^*+\omega_{\vec\pi}(k,0)\psi_{\vec\pi}^*),\cr
\beta_k(t)&=\displaystyle{1\over{2\sqrt{\omega_{\vec\pi}(k,0)}}}
(i\dot\psi_{\vec\pi}+\omega_{\vec\pi}(k,0)\psi_{\vec\pi}),\label{def1}
\end{array}
\end{equation}
and the other possibility is as follows
\begin{equation}
\begin{array}{rl}
\alpha_k(t)&=\displaystyle{1\over{2\sqrt{\omega_{\vec\pi}(k,t)}}}
(i\dot\psi_{\vec\pi}^*+\omega_{\vec\pi}(k,t)\psi_{\vec\pi}^*)
e^{i\int^{t}{} dt'\omega_{\vec\pi}(k,t')},\cr
\beta_k(t)&=\displaystyle{1\over{2\sqrt{\omega_{\vec\pi}(k,t)}}}
(i\dot\psi_{\vec\pi}+\omega_{\vec\pi}(k,t)\psi_{\vec\pi})
e^{-i\int^{t} dt'\omega_{\vec\pi}(k,t')}.\label{def2}
\end{array}
\end{equation}

The former definition (\ref{def1}) relies on time evolutions of
canonical fields, e.g.,
$\vec\pi(k,t)=U(t)\vec\pi(k,0)U^{-1}(t),$ and
Boyanovsky et al. utilizes this type of the number operator for
the numerical analysis. Here $U(t)$ is a time evolution operator.
In the latter definition, the creation and annihilation operators
diagonalize the Hamiltonian at any time $t$ \cite {STB}, and
$a^\dagger a$ defined by these coefficients is often referred
to as {\it adiabatic} number operator.
We will choose the second option, the transformation coefficients
(\ref{def2}) as we did in our previous papers.
By definition, both number operators coincide at $t=0$.

A subtle problem, however, arises in the latter definition.
As noticed in the previous papers \cite {HM98,HM00},
there exists a possibility that
$\omega(k,t)$ becomes imaginary when the amplitude of the
background sigma oscillation exceeds a limit,
$(1 - \varphi_0/v)^2 > (m_\pi/m_\sigma)^2$.
It is highly plausible that the ill-defined imaginary frequency
is an artifact of the wrong choice of variables,
but we do not enter into the problem in this paper.

\section {Single Pion Momentum Distributions and Two Pion Correlations}

The expectation value of the pion number operator is defined by
\begin{equation}
\kakko{n_{\vec\pi}(k,t)}=\mbox{\rm Tr}(a^\dagger_k(t)a_k(t)\rho(0)),
\end{equation}
%%%% insert
where $\rho(0)$ stands for the functional density matrix at $t=0$.
At $t=0$, the initial occupation number is given by
\begin{equation}
\kakko{n_{\vec\pi}(k,0)}={1\over{e^{\beta_0\omega_{\vec\pi}(k,0)}-1}},
\label{initial_n}
\end{equation}
at finite temperature via the first term of right-hand-side
in (\ref{number}), an expected result.
It is easy to obtain the energy in pion sector in terms of a function
$\psi_{\vec\pi}$, which is defined similarly as $\psi_\varphi$ in
(\ref{def_psi}), and it is given by
\begin{equation}
E_{\vec\pi}(k,t)=\left(
\displaystyle{1\over 2}\left|\dot\psi_{\vec\pi}(k,t)\right|^2+
\displaystyle{1\over 2}\omega_{\vec\pi}^2(k,t)
\left|\psi_{\vec\pi}(k,t)\right|^2\right)/
\left(e^{\beta_0\omega_{\vec\pi}(k,0)}-1\right).\label{piE}
\end{equation}

We now proceed to the numerical analysis.
We restrict ourselves into small initial background oscillation 
of the sigma field. We parametrize it, throughout the remaining 
sections, by the parameters 
\begin{equation}
\frac {\chi_0(0)}{v} \equiv \frac {\varphi_0 -v}{v}, 
\end{equation}
which implies a departure from the bottom of the potential 
minimum measured in units of $v$, the vacuum expectation 
value of $\varphi$. 
By solving the coupled differential equations (\ref{eqphi0}) and
(\ref{eqphi}), we compute $\kakko{n_{\vec\pi}(k,t)}$.
In Fig. 1, we present snapshots of the single pion momentum
distributions at time $t=0$, 2, 4, 6, 8, and 10 fm with the initial
amplitude of a background oscillation $\chi_0(0)/v = 0.05$. 
This is the same initial condition as in the previous analysis
ignoring the back reaction \cite {HM00} and the snapshots obtained
there at $t=4$, 7, and 10 fm are presented in Fig. 2 for comparison.
(At $t=0$, $\kakko{n_{\vec\pi}(k,t)}$ trivially vanishes because of 
the initial condition chosen in \cite {HM00}.)
From Fig. 1, one observes the three significant characteristics 
of the time evolution of the pion momentum distribution: 

\vskip 0.3cm 
\noindent
(1) At momenta lower than $\sim 100$ MeV, the number density 
grows first and then decreases, and seems to undergo damped oscillations. 
The suppression of number density at low momenta seems to indicate 
that the back reaction due to particle creations are efficient at 
low momenta. It may be a bad news for DCC because it may suppress 
the formation of large coherent domains. We will 
make some comments later.

\vskip 0.3cm
\noindent
(2) At around the resonance energy $\sim 200$ MeV, a peak starts 
to develop at $t \simeq 4$ fm and continues to grow as time goes by. 
The series of snapshots in Fig. 1 looks like a process that the peak 
grows by absorbing the ambient pions in low momentum region which 
was originally provided, as seen by comparing with Fig. 2, by the 
background oscillation in an early stage of evolution.

\vskip 0.3cm
\noindent
(3) At momenta larger than $\sim 300$ MeV, there is virtually 
no change in the initial thermal distribution. 

\vskip 0.3cm

The peak height of pion number distribution at resonance without 
the back reaction reaches 0.28 at $t=10$ fm in Fig. 2. 
The corresponding peak height with back reaction is 0.38 at $t=10$ fm 
as shown in Fig. 1. The peak height with the back reaction is higher 
but it is mainly due to the fact that the initial occupation number 
(\ref{initial_n}) is nonzero. 
If we estimate the net increase of height due to evolution just by 
subtracting the initial thermal distribution, we obtain 0.25. 
It amounts to about 90\% of the peak height without the back reaction.

We also note, by comparing Fig. 1 with Fig. 2, that the peak 
position moves to lower momentum region when the back reaction is 
taken into account. 
It can be understood by the following consideration. 
%

%%%% insert
If we ignore back reactions, the equation of motion of pion fields 
reduces to the Mathieu equation
\begin{equation}
\left({d^2\over{dz^2}}+A-2q\cos(2z)\right)\pi_k=0,
\end{equation}
where $A=4(k^2+m_\pi^2)/m_\sigma^2$ and 
$q=4\lambda v\chi_0(0)/m_\sigma^2$ \cite{HM00}. 
Here we used a dimensionless time variable defined by $z=m_\sigma t/2$.
It is well known that the equation admits unstable solutions in a wide 
range of parameters, in particular at $A=n^2$ for small $q$ which is 
relevant to our case, where $n$ denotes an arbitrary integer \cite{mathieu}. 
Using the effective pion mass corrected by back reactions 
(\ref{mass}) the peak position is roughly estimated as
\begin{equation}
k_{peak}\sim \sqrt{{m_\sigma^2\over 4}-m_\pi^2\cdot{v_0\over v}-
\lambda\chi_0^2-3\lambda\kakko{\pi^2}},\label{pposition}
\end{equation}
for the first resonance band $A=1$. 
%%%

The back reaction affects $k_{peak}$ in two opposite ways.
The amplitude of the oscillation obviously damps by back reactions 
and it makes the third term in the square root smaller in magnitude
at $t \ne$ 0, and hence $k_{peak}$ larger.  
On the other hand, the last term, which was absent in a case of
no back reaction, tends to let $k_{peak}$ be smaller. 
A simple estimate shows that the latter effect wins. 

To confirm this interpretation, we run the computations with 
four different initial conditions in the region $\chi_0/v\in [0.04,0.10]$.
In Fig. 3, we plot the pion distributions at $t=20$ fm. 
One notices that the peak position moves toward a lower momentum region
by increasing the amplitudes of background sigma oscillation
in consistent with our interpretation. It is also notable that the 
peak height rapidly increases as $\chi_0/v$ gets larger. It clearly 
shows a hint for experiments.

The behavior of damping of the background oscillation is shown 
in Fig. 4.
This and the resultant increase of the effective mass cause a shift of 
the peak position to a lower momentum, as we just argued.  
In Fig. 5, we plot the time evolution of the energy of pion sector,
$E_{\pi}$. It is defined as the summation over $k$ in (\ref{piE}). 
It increases while the background sigma oscillation damps with 
characteristic time scale of $\sim 10$ fm, and becomes stationary 
at about $t \sim 15$ fm when the background sigma oscillations die 
away. The pion production effectively terminates at this point.

The momentum distribution for sigma fluctuations is shown in
Fig. 6. It represents a sharp contrast with the behavior of pion 
distributions, having no evolution until $t=30$ fm. 
The first resonance is expected at zero momentum, but it is not 
visible. It may be due to cancellation between the resonance 
enhancement and the damping due to back reactions at low momenta.

We have observed that the time scale of the energy dissipation 
is $\sim 10$ fm.
The time scale can roughly be understood by the following arguments.
For simplicity, we consider a system with pion and background fields 
only. Suppose that the background fields oscillate harmonically. 
Using the virial theorem, the energy conservation reads
\begin{equation}
m_\sigma^2\tilde\chi_0(0)^2+\epsilon_\pi(0)=
m_\sigma^2\tilde\chi_0(z)^2+\epsilon_\pi(0)e^{2\mu z},
\label{econservation}
\end{equation}
%%% insert
since the pion field $\pi_k(z)$ is expressed by using the critical
exponent as $\pi_k(0)e^{\mu z}$ in a resonance band.
The pion energy density $\epsilon_\pi$ is proportional 
to the square of the pion field $\pi_k(z)$.
$\tilde\chi_0$ implies the amplitude of background oscillation, and
$\mu$ is the critical exponent of the Mathieu function \cite{mathieu}.
As mentioned above, 
%%%
the equation of motion of pion is described by the 
Mathieu equation in this simple system and the energy of pion is dominated 
by the modes in the resonance band in a good approximation.

In our analysis, a parametric resonance occurs in a narrow resonance 
band and $\mu$ is given 
%%% insert 
by $\mu=q/2$ \cite{mathieu}, that is
%%%
\begin{equation}
\mu={2\lambda v \tilde\chi\over m_\sigma^2}=
{2\lambda v\sqrt{\epsilon_{0} -\epsilon_\pi(z)}\over m_\sigma^3},
\end{equation}
where $\epsilon_{0}$ is the initial total energy density of 
the system considered.
Because the time dependence of pion energy at any time $z$ is
expressed by
$\epsilon_\pi(z)=\epsilon_\pi(0)e^{2\mu z}$,
one obtain the following equation
\begin{equation}
{d\epsilon_\pi(z)\over{dz}}=2\mu\epsilon_\pi(z)\label{energy_eq}
\end{equation}
in the region where the time derivative of $\mu$ can be
neglected. Practically, it means the narrow resonance region.
The solution is given by
\begin{equation}
\epsilon_\pi(z)=\epsilon_{0} \left(1-\tanh^2\left[
-{2\lambda v\sqrt{\epsilon_{0}}\over m_\sigma^3}z+
\tanh^{-1}\sqrt{1-{\epsilon_\pi(0)\over \epsilon_{0}}}\right]\right),
\end{equation}
where $z\in [0, z_{end}]$. 
The time $z_{end}$ is a dimensionless cut off time beyond which 
the equation (\ref{energy_eq}) is no longer valid.
One can see that almost all energy of background field is transferred to
pions by the time given by 
\begin{equation}
z_{end}={m_\sigma^3\over{2\lambda v\sqrt{\epsilon_{0}}}} 
\tanh^{-1}\sqrt{1-{\epsilon_\pi(0)\over \epsilon_{0}}}.
\end{equation}
The energy density $\epsilon_{0}$ is determined, once the initial 
amplitude of sigma oscillation is given. 
With parameters used in our numerical calculation, $\chi_0/v=0.05$, 
$\epsilon_{0} \simeq 8.67 \times 10^{6}$ MeV${}^4$. 
Using the parameters given in (\ref{parameters}), $t_{end}$ is 
estimated to be $\sim 8$ fm. 
Note that this value is under-estimated because
$\mu$ is treated as a constant in the method we used.
Since $\mu$ decreases with time, the energy transfer needs
longer time to be completed.
Nevertheless, it is notable that the treatment gives a correct order
of magnitude estimation of $z_{end}$.

Two point correlation function $C(t,r)$ is similarly obtained by means
of the density matrix, i.e.,
\begin{equation}
\begin{array}{rl}
C(t,r)&=\kakko{\vec{\pi}(t,0)\vec{\pi}(t,r)}
\equiv Tr(\vec{\pi}(t,0)\vec{\pi}(t,r) \rho(0))
\cr
&=\displaystyle{1\over{\Omega^{1/3}}}\sum_k\displaystyle{k^2\over{2\pi}}
j_0(kr)\left|\psi_\pi\right|^2\coth\left(
\displaystyle{\beta_0\omega_{\vec\pi}(k,0)\over 2}\right),
\end{array}
\end{equation}
where $j_{0}(x) = \sin{x}/x$ denotes the spherical Bessel function 
of order 0.
In Fig. 7, we plot two point pion correlation functions at 
$t=5$, 10, and 20 fm.
At $t=10$ when the evolution effectively ends, 
the correlation length $\xi$ is given by $\xi \simeq 1.3$ fm.

In search for the possibility of correlation length growth, 
we run the computations with varying initial conditions. 
The correlation length increases for larger initial background
oscillations, but not so much, as we observe in Table.1.

%%%%%%%%% Table 1 %%%%%%%%%%
\begin{table}
\caption{Correlation length $\xi$ of pions at $t=20$ fm for 
varying initial background amplitude $\varphi_0/v$'s.}

\vskip 0.7cm
\begin{center}
\begin{tabular}{ccccccc}
\hline\hline
$\chi_0(0)/v$&0.05&0.06&0.07&0.08&0.09&0.10\cr
$\xi$[fm]&1.27&1.36&1.43&1.50&1.61&1.68\cr
\hline\hline
\end{tabular}
\end{center}
\end{table}

Apparently, this result is in disagreement with the conclusion 
reached by Kaiser who found the large domain formation due to 
the parametric resonance mechanism \cite {kaiser}. 
Our simulation indicates that the back reaction primarily affects 
low momentum components and it is unlikely that the mechanism 
itself leads to large domain formation.

We note, however, that there are two significant differences between 
our and his calculations: 

\vskip 0.3cm 
\noindent
(i) The initial amplitude of background sigma oscillation is large, 
$\chi_{0}/v \sim$ 1 in Kaiser's calculations, whereas it is only 
up to 0.1 in our present analysis. 

\vskip 0.3cm 
\noindent
(ii) The quantum back reaction is fully taken into account in each 
step of evolution of the system under the Hartree-Fock approximation 
in our calculation. But it is ignored in Kaiser's analysis apart from 
that the effect is evaluated to determine the time at which the 
simulation has to stop. 

\vskip 0.3cm 

At this stage, we feel it difficult to judge whether or not 
large coherent domains are expected to form by the parametric 
resonance mechanism. It may well be the case that it depends 
upon the initial conditions. 
To really address the issue, we have to run a simulation with large 
initial amplitudes with full taking account of quantum back reactions. 
To carry it out, we need a formulation which is free from the 
instability problem as we mentioned at the end of Sec. 2.

\section {Concluding Remarks}
In this paper we have investigated features of quantum back reactions 
due to particle production by the parametric resonance mechanism in 
an environment of nonequilibrium chiral phase transition.
We have calculated the single pion momentum distributions and 
two pion correlations under the initial small amplitude sigma 
oscillation around a potential minimum and with initial thermal 
equilibrium assumption.
We relied on the formalism developed by Boyanovsky et al. 
to take account of back reactions under the Hartree-Fock 
approximation based on the Schr\"odinger picture formulation 
of quantum field theory. Remarkably, we observed that the 
resonance peak survives under the back reaction even in such a strongly 
coupled linear sigma model.

If the setting of small initial background oscillation is relevant 
in physical situation in high-energy hadronic collisions, a sharp peak 
in pion momentum distributions can be a clear signal for the 
parametric resonance mechanism, one of the candidate mechanism 
for DCC. 

The potential possibility of large domain formation, the key issue in 
DCC, is not fully explored and hence unanswered in the present analysis. 
We hope that we can return to this problem by having a 
formalism without instability problem in the near future.
  
\section {Acknowledgment}

We thank Robert Brandenberger for bringing \cite {brand} to our attention. 
This work was supported in part by the Grant-in-Aid for Scientific
Research No. 12640285, Ministry of Education, Culture, Sports, 
Science and Technology of Japan. The research of H.H. was partly supported
by Kitasato University Research Grant for Young Researchers.

\newpage

%%%%%%%%%  References  %%%%%%%%

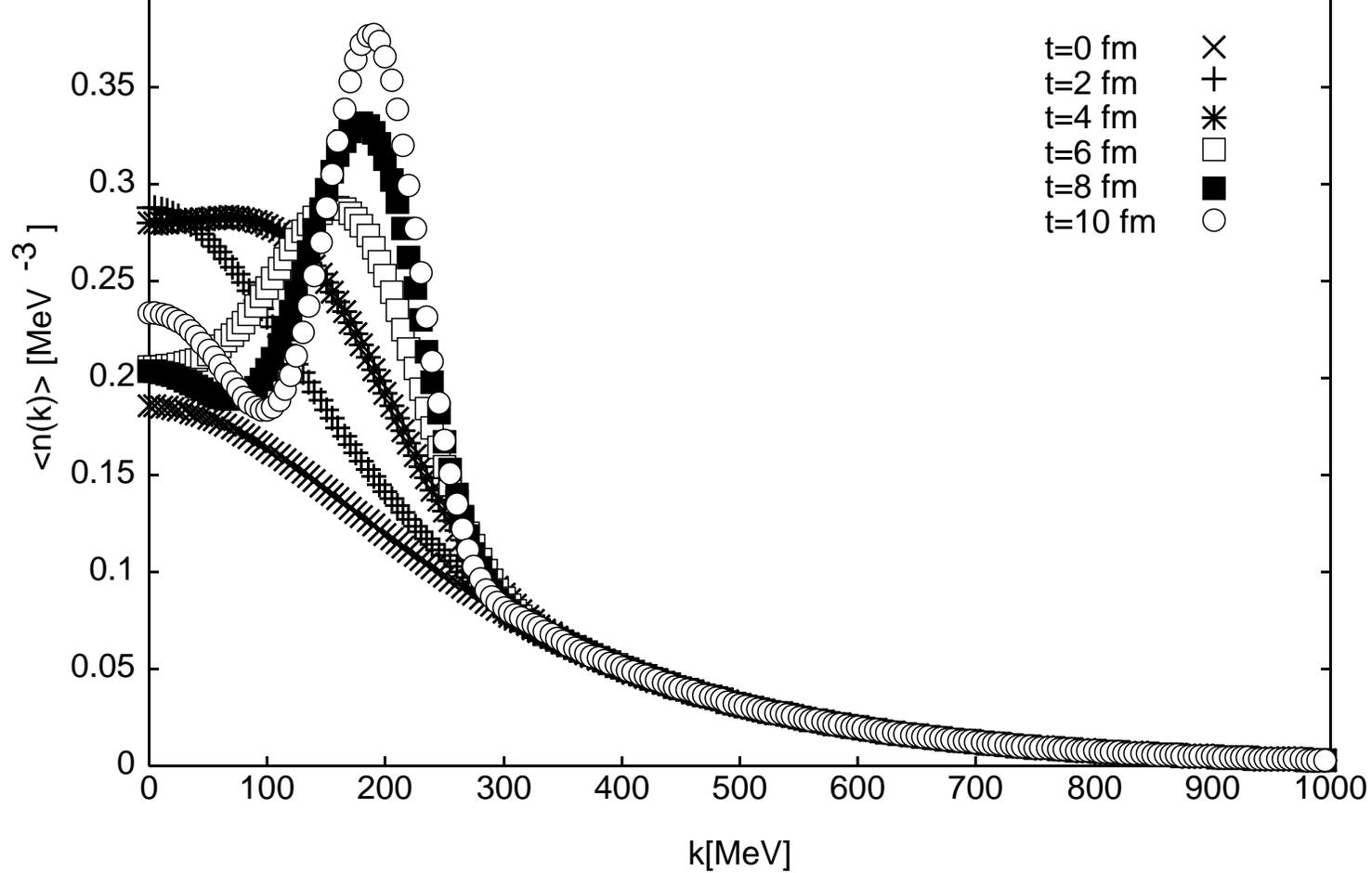
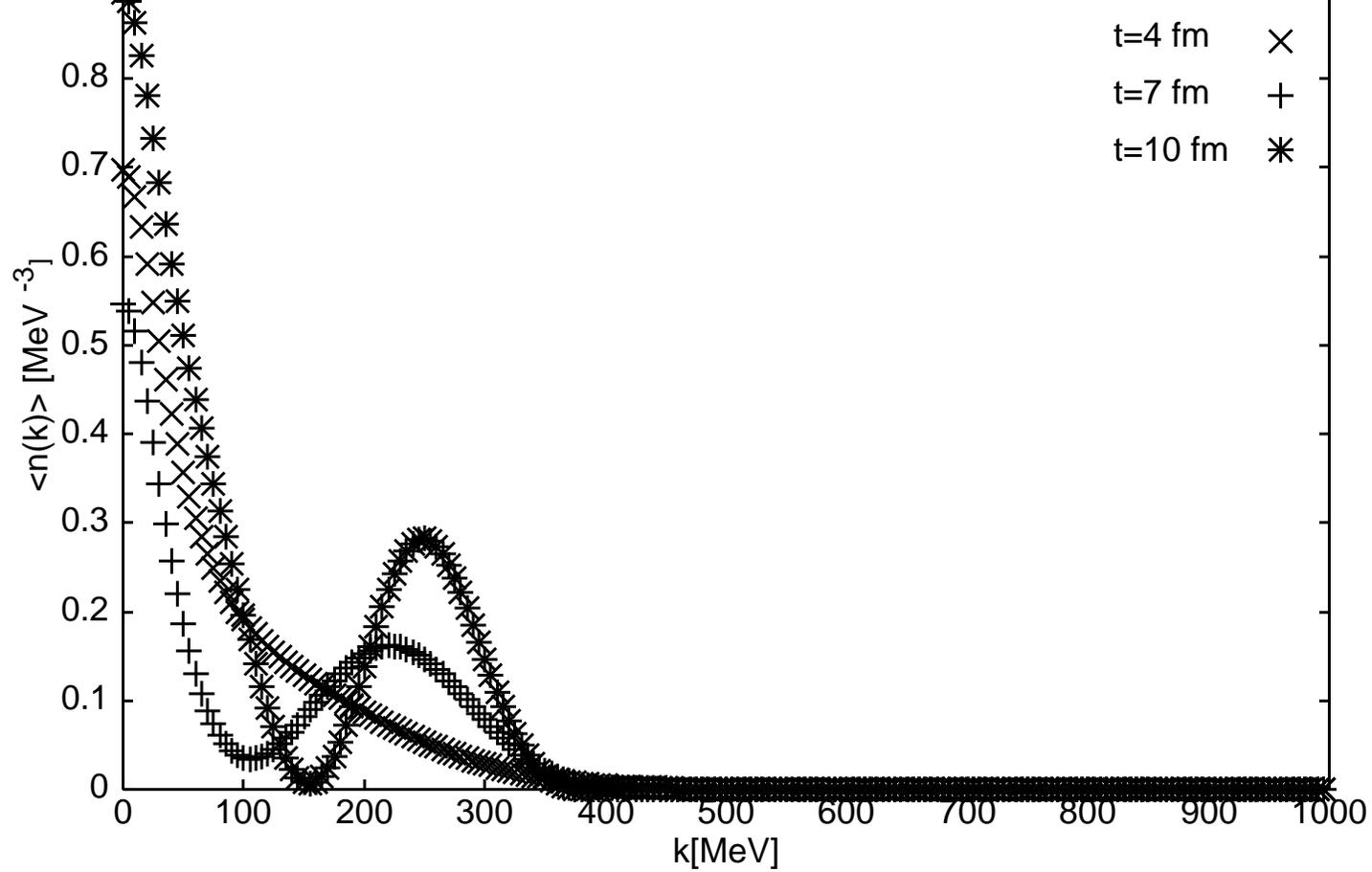
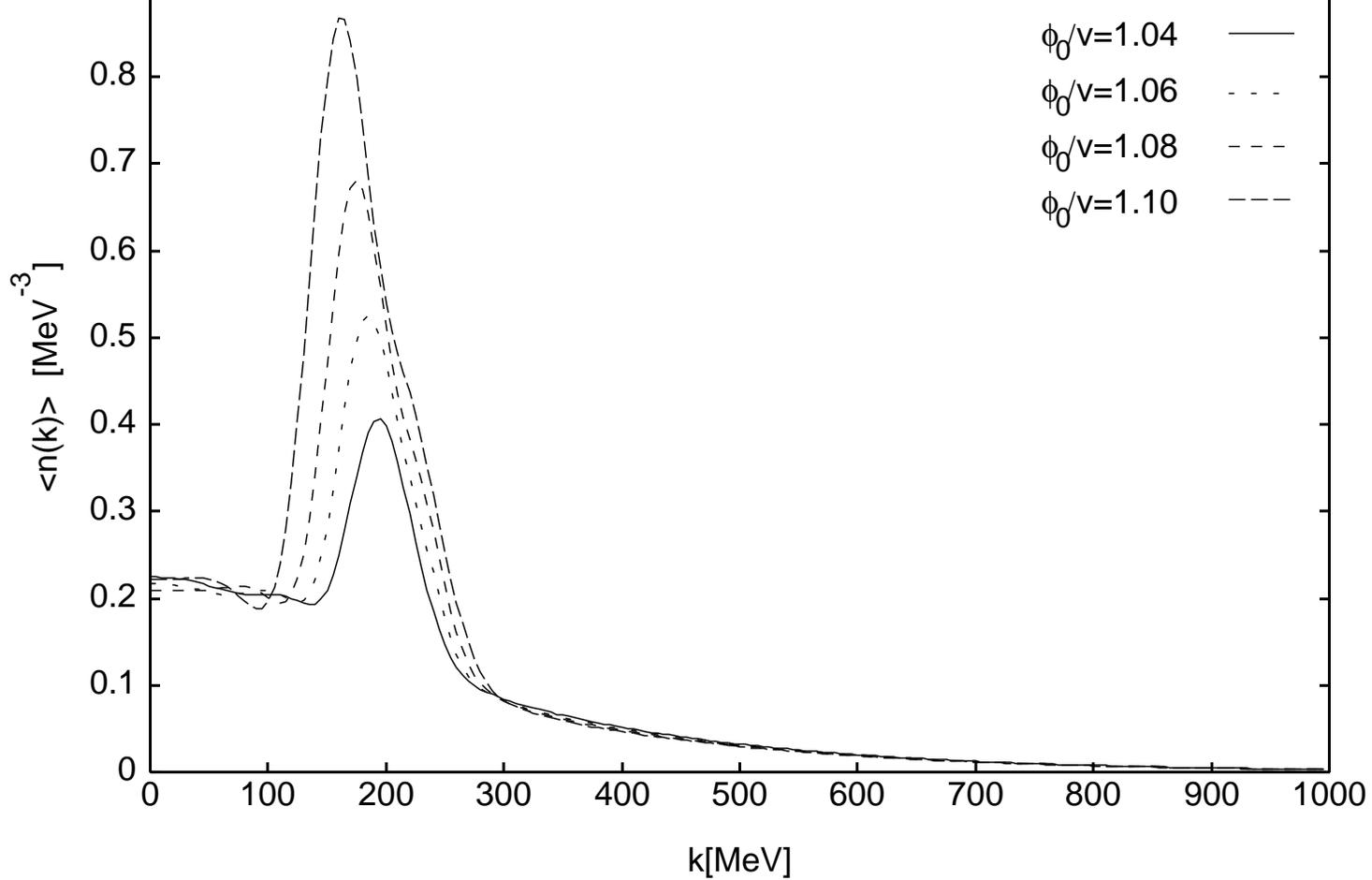
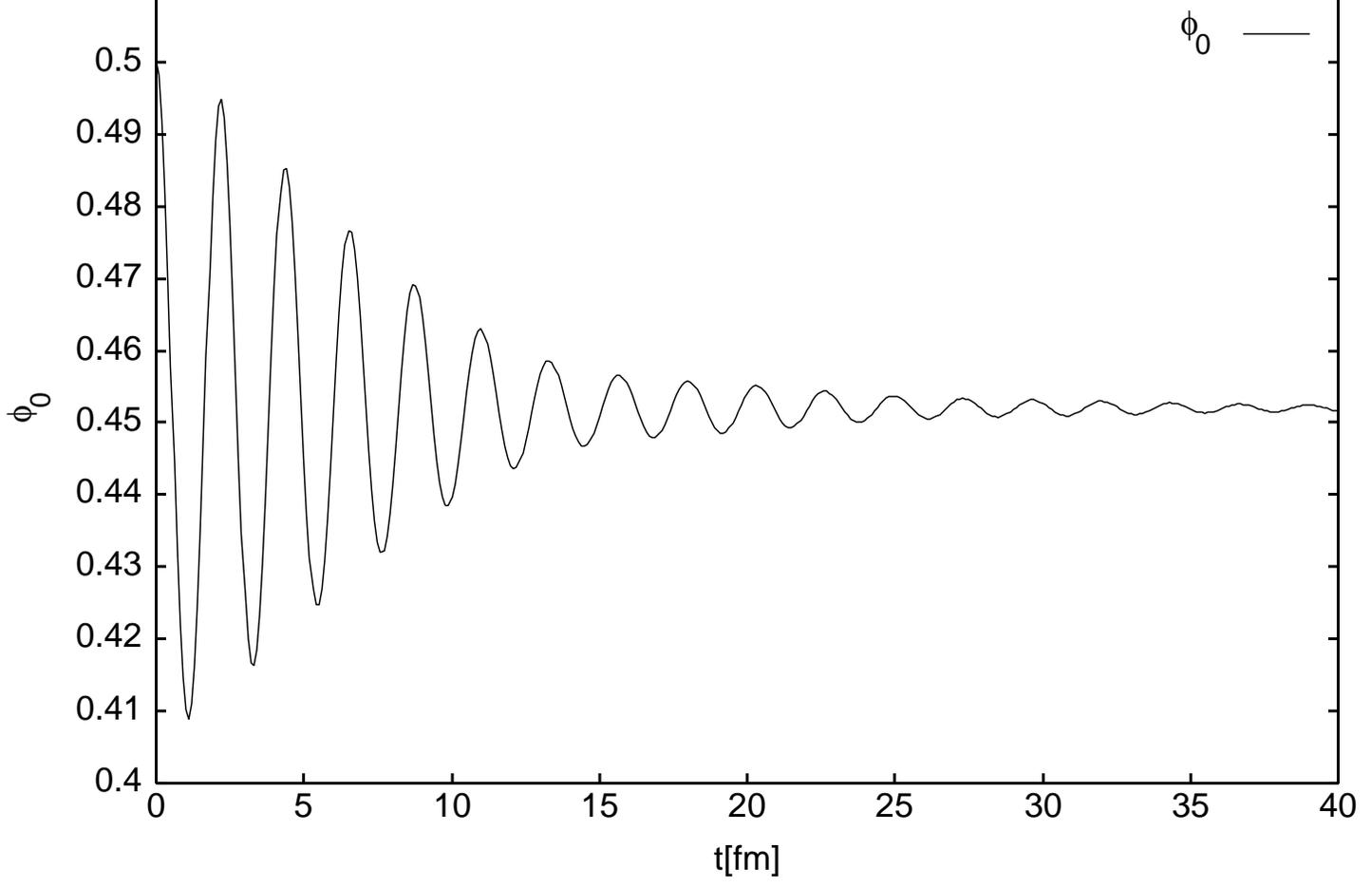
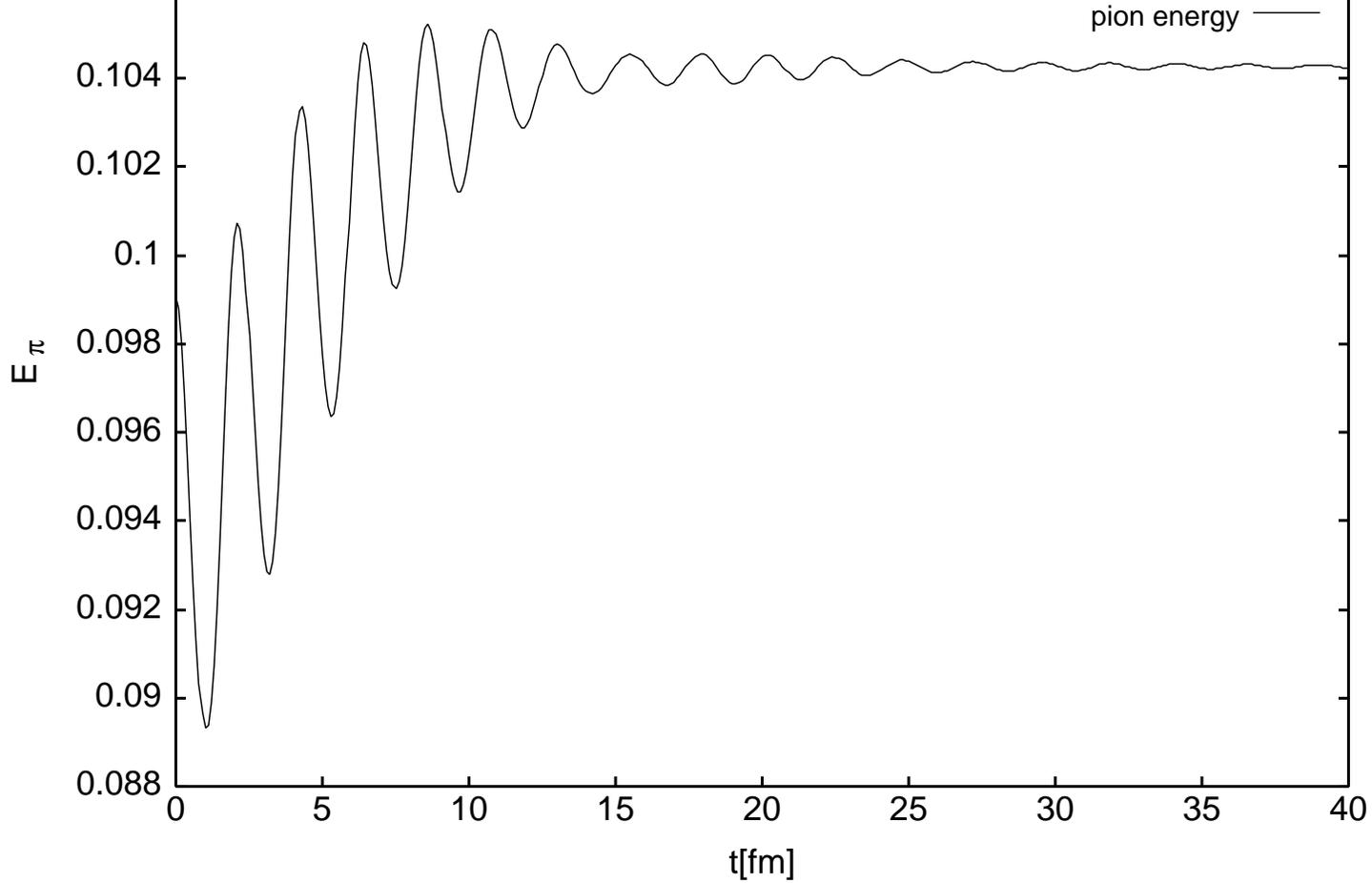
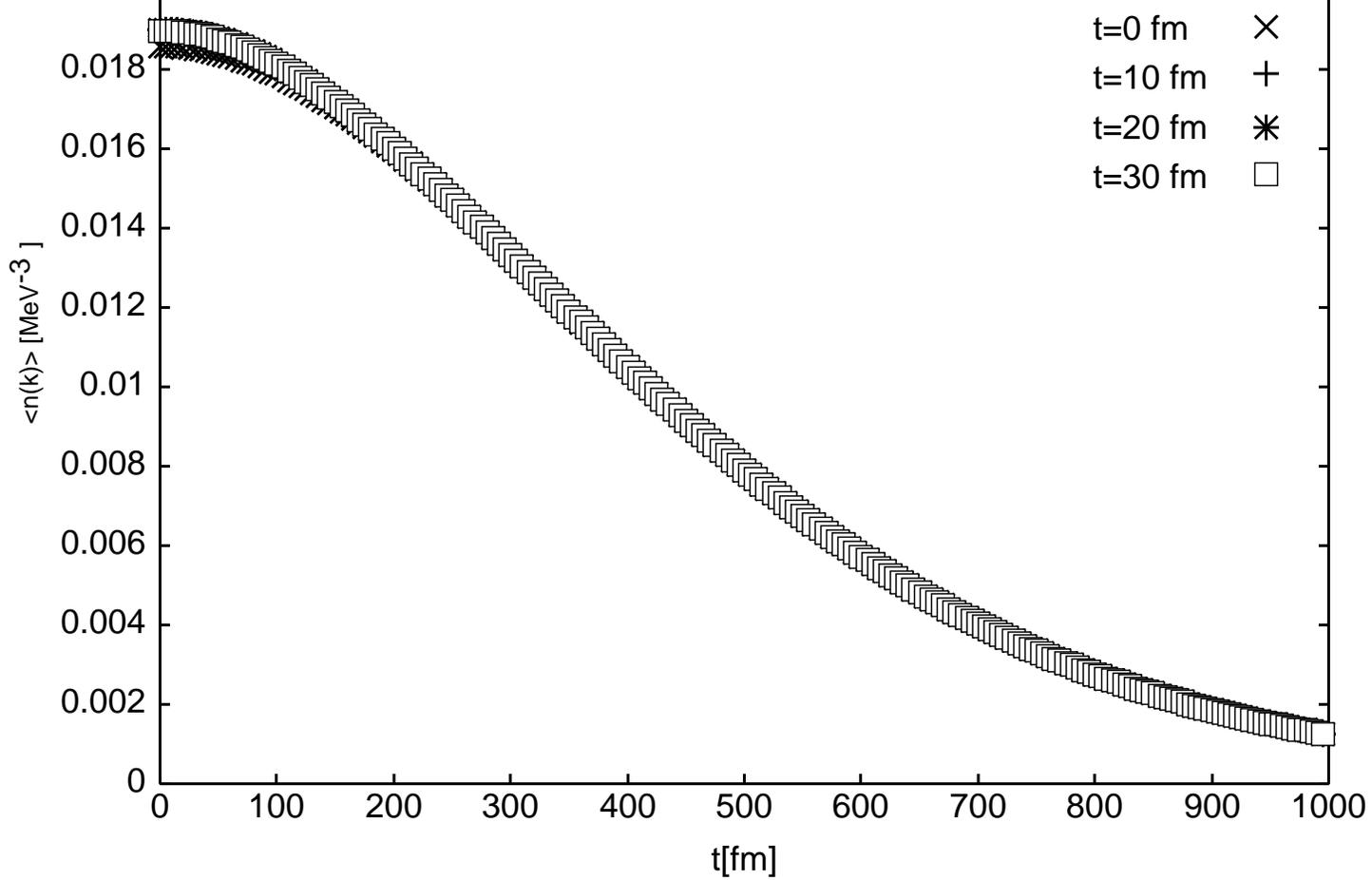
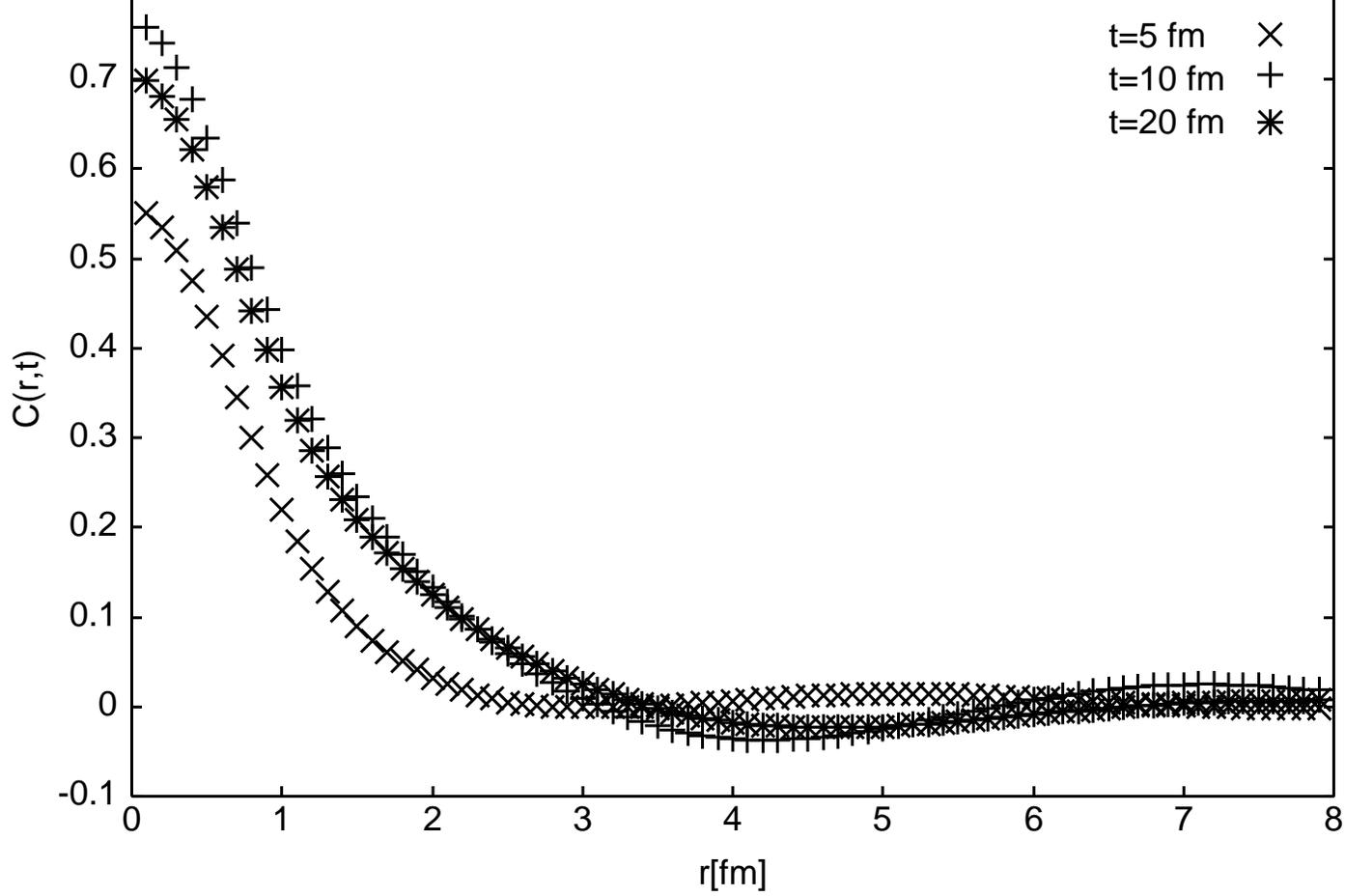
\end{document}